\documentclass[a4paper,11pt]{article}

\usepackage{jheppub} 

\usepackage[T1]{fontenc} 

\title{\boldmath Energy-momentum tensor of special relativistic fluids and Connection of relativistic and non-relativistic fluids in the new scaling}

\author[1]{M. Moeen,\note{Corresponding author.}}

\affiliation[1]{Kosar University of Bojnord, Department of Sciences, Iran, Bojnord}

\emailAdd{Dr.moeen@kub.ac.ir}

\abstract{
In this paper, the relativistic fluids with the negligible magnetic field in the flat metric are studied. The iso-dimension scale is introduced, in this scale, all similar components of a variable have the same dimension. Also, the components of shear, bulk and heat flux tensors and the components of energy-momentum tensors are calculated in cylindrical and spherical coordinate systems, all these components are derived in the iso-dimension scale. 
The non-relativistic shear, bulk, and heat flux tensors and the components of energy-momentum tensor are derived in the limit of the relativistic fluids. Also, the connection of relativistic and non-relativistic fluids are seen. So, some distinctions of the relativistic and non-relativistic fluids are displayed, for example, in the relativistic fluids, the time derivative of velocity is created the shear stress tensor and time derivative of velocity and temperature is created the heat flux energy-momentum tensor, etc. }

\begin{document} 
\maketitle  
\flushbottom
\section{Introduction}\label{s2} 
In astrophysics, the relativistic fluids are used to study black holes, accretion disks, neutron stars, white dwarf, X-ray binaries, etc. 
The influences of the relativistic viscosity and heat flux were studied by many authors. 
In \cite{shi} a simple form of relativistic shear viscosity in the Cartesian coordinate system for study the neutron stars was used, so, bulk viscosity and heat conduction were neglected in the transformation processes. In \cite{lah} the influences of simple and relativistic shear viscosity was applied for study the shape of thick accretion disks, this shear viscosity was created because of coupling between shear viscosity and curvature. In \cite{du} the influences of shear viscosity in the stability of neutron stars were studied. \cite{laa} used the non-relativistic heat conduction in the study of relativistic stars . In \cite{mo18a}, the influences of relativistic shear and bulk viscosity in the thermodynamics quantities of accretion disks in the Kerr metric was studied . In some studies, the causal viscosity was used to study of relativistic accretion disks around the rotating black hole (\cite{ta}, \cite{gp}, etc.) 
In a wide range of papers, the relativistic heat flux, shear, and bulk viscosity and shear stress tensors were derived. In \cite{mo17} the components of shear and bulk tensor of relativistic fluids in the Kerr metric was calculated. The components of heating tensor and viscous stress tensor with the zero bulk viscosity were derived in the Cartesian coordinate system in \cite{mo18}. The non-ideal dynamics of the relativistic fluids has been studied by \cite{mu}. In this paper, the components of shear and bulk tensor and heat flux in some cases were derived. \cite{hs} have studied the dissipative dynamics of relativistic fluids in the equatorial plane. The components of shear and bulk tensors in the Kerr metric in the equatorial plane were obtained in \cite{mo17}. 
The relativistic fluids with the negligible magnetic fields in the flat metric are studied in this paper. In most papers, the ordinary scaling $\hbar=c=G=1$ were used, we don't use these scales. We introduce the iso-dimension scale, in this scale all components of the metric are dimensionless, also all components of the same quantities have the same dimension, and influences of the speed of the light are seen in this scale. All calculations are done in this scale. 
In the relativistic fluids, the energy-momentum tensor is important, because this tensor shows the quantities of energy and momentum of fluids. The energy-momentum tensor of fluids with the ignorable magnetic fields includes three types of energy-momentum tensors, which are perfect fluid, shear stress viscosity , and heat flux energy-momentum tensors. 
The components of a perfect fluid, shear, bulk, shear stress energy-momentum, heat flux, and heat flux energy-momentum tensors are calculated in cylindrical and spherical coordinate systems. 
The iso-dimension scale is enabled us to calculate the components of a perfect fluid, shear, bulk, shear stress energy-momentum, heat flux , and heat flux energy-momentum tensors of non-relativistic fluids as the limit of these quantities in the relativistic fluids. So, some of the differences of non-relativistic and relativistic fluids are seen. For example, we see that in the relativistic fluids, the heat flux is created in the constant temperature, etc. 
The plan of this paper is as follows: In section \ref{s2} basic conservation equations are shown. Scale and metric are in section \ref{s3}. In this paper two coordinate systems are used, the components of metric in these coordinate systems are in section \ref{s4}. Four-velocity in two of these coordinate systems is displayed in section \ref{s5}. The perfect fluids energy-momentum tensor is derived in section \ref{s6}. The relations of all components of shear and bulk tensors are derived in section \ref{s7} . The shear stress viscosity energy-momentum tensor is calculated in section \ref{s8}. the relativistic heat flux and energy-momentum of heat flux are in sections \ref{s9} and \ref{s10}. The energy density of relativistic fluids is in section \ref{s11}. The components of the energy-momentum of non-relativistic fluids are given in section \ref{s12}. Summary and conclusion are in section \ref{s13}. 
\section{Basic conservation equations}\label{s2} 
The basic conservation equations of relativistic fluids are 
\begin{equation}\label{1} 
\rho u^{\nu}_{;\nu}=0,\qquad T^{\mu\nu}_{;\nu}=0. 
\end{equation} 
The first equation is the mass conservation equation and the second is the energy-momentum conservation equation. Where, $\rho$ is the energy density, $ u^{\nu} $ is four-velocity and $T^{\mu\nu}$ is the energy-momentum tensor. 
\subsection{Relativistic Energy- momentum tensor}
The relativistic energy-momentum tensor for non-magnetic viscous fluids with heat flux is given by \cite{pl} 
\begin{equation}\label{2} 
T^{\mu\nu}=T^{\mu\nu}_{Pf}+T^{\mu\nu}_{visoc}+T^{\mu\nu}_{heat}, 
\end{equation} 
where, respectively are perfect fluid, shear stress viscosity, and heat flux energy-momentum tensors. The components of energy-momentum tensor include various information about the energy and momentum of fluids (\cite{sc}). So, the energy density is given by $T^{00}$ ($0$ is for the time like components); Also, the energy flux passes from the surface with the constant $x^{i}$ ($i$ is for space like components) is shown by $T^{0i}$ and the flux of the $i$ components of momentum passes from the surface with the constant $x^{j}$ is given by $T^{ij}$ ($i$ and $j$ are the space like components). 
\subsubsection{ Energy momentum tensor of perfect fluid} 
The perfect fluid energy momentum tensor is 
\begin{equation}\label{3} 
T^{\mu\nu}_{Pf}=\rho u^{\mu}u^{\nu}+pg^{\mu\nu}, 
\end{equation} 
where, $P$ is the pressure and $ g^{\mu\nu} $ are components of metric.
\subsubsection{Energy momentum tensor of shear stress viscosity } 
The shear stress viscosity energy momentum tensor is 
\begin{equation}\label{4} 
T^{\mu\nu}_{vis}=S^{\mu\nu}+B^{\mu\nu}, 
\end{equation} 
$S^{\mu\nu}$ is the relativistic shear viscosity and $B^{\mu\nu}$ is the relativistic bulk viscosity, which are shown as 
\begin{equation}\label{5} 
S^{\mu\nu}=-2\lambda\sigma^{\mu\nu},\qquad 
B^{\mu\nu}=-\zeta b^{\mu\nu}, 
\end{equation} 
where, $\lambda$ is the coefficient of the dynamical viscosity, $\sigma^{\mu\nu}$ is the shear tensor, $\zeta$ is the coefficient of the bulk viscosity, $b^{\mu\nu}$ is the bulk tensor. The relativistic bulk and shear tensors are given by
\begin{equation}\label{6} 
b^{\mu\nu}=\Theta h^{\mu\nu}, \qquad 
\sigma^{\mu\nu}=g^{\mu\alpha}g^{\nu\beta}\sigma_{\alpha\beta},
\end{equation} 
where, the shear rate, $\sigma_{\alpha\beta}$, is derived as (\cite{mt}) 
\begin{equation}\label{7} 
\sigma_{\alpha\beta}=\frac{1}{2}(u_{\alpha;\beta}+u_{\beta;\alpha}+\frac{1}{c^{2}}(a_{\alpha}u_{\beta}+a_{\beta}u_{\alpha}))-\frac{1}{3}\Theta 
h_{\alpha\beta}. 
\end{equation} 
Also, the covariant derivative of $u^{\nu}$ and $u_{\nu}$, the projection tensor ($h^{\mu\nu}$), the expansion of fluid world line ($\Theta$), four acceleration($a^{\mu}$) are written as 
\begin{eqnarray}\label{8} 
&& u^{\nu}_{;\gamma}=\partial_{\gamma}u^{\nu}+\Gamma^{\nu}_{\gamma\lambda}u^{\lambda},\qquad u_{\nu;\gamma}=\partial_{\gamma}u_{\nu}-\Gamma^{\lambda}_{\nu\gamma}u_{\lambda}, 
\nonumber\\&& h^{\mu\nu}=h^{\nu\mu}=g^{\mu\nu}+\frac{u^{\mu}u^{\nu}}{c^{2}},\quad 
\Theta=u^{\gamma}_{;\gamma}=\partial_{\gamma}u^{\gamma} +\Gamma_{\gamma\nu}^{\nu}u^{\gamma},\nonumber\\ && a^{\mu}=u^{\mu}_{;\gamma}u^{\gamma} . 
\end{eqnarray} 
where, $\Gamma^{\alpha}_{\mu\nu}=\Gamma^{\alpha}_{\nu\mu}=\frac{1}{2}g^{\alpha\rho}(\partial_{\mu}g_{\nu\rho}+\partial_{\nu}g_{\rho\mu}-\partial_{\rho}g_{\mu\nu})$ are Christoffel symbols. 
Also, shear tensor is derived as 
\begin{eqnarray}\label{9} 
&&\sigma^{\mu\nu}=g^{\mu\alpha}g^{\nu\beta}[u_{\alpha;\beta}+u_{\beta;\alpha}+\frac{1}{c^{2}}(a_{\alpha}u_{\beta}+a_{\beta}u_{\alpha})-\frac{1}{3}\Theta 
h_{\alpha\beta}] 
\nonumber\\ && \qquad =\frac{1}{2}(g^{\nu\beta}u^{\mu}_{;\beta}+g^{\mu\alpha}u^{\nu}_{;\alpha}+\frac{1}{c^{2}}(a^{\mu}u^{\nu}+a^{\nu}u^{\mu}))-\frac{1}{3}\Theta h^{\mu\nu}. \nonumber\\
\end{eqnarray} 
\subsubsection{ Energy momentum tensor of heat flux } 
The heat flux energy momentum tensor is given by 
\begin{equation}\label{10} 
T^{\mu\nu}_{heat}=\frac{1}{c^2}(q^{\mu}u^{\nu}+q^{\nu}u^{\mu}), 
\end{equation} 
where, $q^{\mu}$ is heat flux. So, the relativistic heat flux is shown as (\cite{ek} and \cite{mi}) 
\begin{equation}\label{11} 
q^{\mu}=-\kappa (\frac{\partial T}{\partial x^{\nu}}+\frac{T a_{\nu}}{c^{2}}), 
\end{equation} 
where, $ T $ is temperature and $ \kappa $ is thermal conductivity. 
\section{Scale and Metric}\label{s3} 
One of the ordinary scale is used in the study of relativistic fluids is $\hbar=c=G=1$. The benefit of this scale is that all variables are dimensionless, so, the equations in this scale are convenient and straightway, also, the amounts of these constant are not seen in equations. If we want to establish the connection between relativistic and non-relativistic fluids we must retain importance of $c$ in the equations. So, in this paper a new scaling is used which named iso-dimension scale (IDS). In this scale all components of metric are dimensionless. Also, different components of a typical tensor have a same dimension. 
The relativistic fluids are seen in astrophysics, cosmology, nuclear physics, etc. We study the relativistic fluids in the region which the influences of magnetic field and gravitation are negligible, So, in this condition flat metric for space-time metric are used. For example in the astrophysical fluids the flat space-time relativistic fluids can be used around the proto stars, young stars, far from black holes ($r>>\frac{GM}{c^2}$, $M$ is the central mass), etc. 
\section{Coordinate systems}\label{s4} 
We study the special relativistic fluids in two coordinate systems (cylindrical and spherical coordinate systems).
\subsection{Cylindrical coordinate system} 
In the cylindrical coordinate system, the relativistic flat metric (Minkowski space-time) is given as
\begin{equation}\label{12} 
ds^{2}=-c^{2}d\tau^{2}=-c^{2}dt^{2}+dr^{2}+r^{2}d\theta^{2}+dz^{2}, 
\end{equation} 
where, $c$ is speed of light, $\tau$ is the proper time, $t, r, \theta $ and $z$ are coordinate variables. $g_{\mu\nu}$ and $g^{\mu\nu}$ are diagonal tensors, so, the components of metric, and its inverse, are shown
\begin{equation}\label{13} 
g_{tt}=-c^2,\qquad g_{rr}=1,\qquad g_{\theta\theta}=r^{2},\qquad g_{zz}=1. 
\end{equation} 
\begin{equation}\label{14} 
g^{tt}=-\frac{1}{c^2},\qquad g^{rr}=1,\qquad g^{\theta\theta}=\frac{1}{r^{2}},\qquad g^{zz}=1, 
\end{equation} 
$g_{rr},g^{rr},g_{zz}, g^{zz}$ are dimensionless, but $g_{tt},g^{tt},g_{\theta\theta},g^{\theta\theta}$ have different dimensions. 
In this paper, a simple and useful method to iso-dimensioning of metric components is used. We use a constant $l$ in $g_{\theta\theta},g^{\theta\theta}$ components, so, $\psi=l\theta$, also $\chi=x^{0}=tc$ is used. Therefore, the metric is converted to the iso-dimensional coordinate 
\begin{equation}\label{15} 
ds^{2}=-c^{2}d\tau^{2}=-d\chi^{2}+dr^{2}+R^{2}d\psi^{2}+dz^{2}, 
\end{equation} 
where, $R=\frac{r}{l}$ is non-dimensional variable, $l$ is arbitrary, finite constant length and $\psi=l\theta$. Also, the non-dimensional components of metric, $g_{\mu\nu}$, and its inverse, $g^{\mu\nu}$ are 
\begin{eqnarray}\label{16} 
&&g_{00}=-1,\qquad g_{rr}=1,\qquad g_{\psi\psi}=R^{2},\qquad g_{zz}=1, 
\nonumber\\ &&g^{00}=-1,\qquad g^{rr}=1,\qquad g^{\psi\psi}=\frac{1}{R^{2}},\qquad g^{zz}=1.\nonumber\\ 
\end{eqnarray}
\subsection{Spherical coordinate system} 
The flat metric in the spherical coordinate system is as fallows 
\begin{equation}\label{17} 
ds^{2}=-c^{2}d\tau^{2}=-c^{2}dt^{2}+dr^{2}+r^{2}d\theta^{2}+r^{2}sin^{2}\theta d\phi^{2}, 
\end{equation} 
where, $t, r, \theta $ and $\phi$ are coordinate variables. The components of diagonal metric $g_{\mu\nu}$, and its inverse, $g^{\mu\nu}$ in the spherical coordinate are given as 
\begin{equation}\label{18} 
g_{tt}=-c^2,\qquad g_{rr}=1,\qquad g_{\theta\theta}=r^{2},\qquad g_{\phi\phi}=r^{2}sin^{2}\theta. 
\end{equation} 
\begin{equation}\label{19} 
g^{tt}=-\frac{1}{c^2},\qquad g^{rr}=1,\qquad g^{\theta\theta}=\frac{1}{r^{2}},\qquad g^{\phi\phi}=\frac{1}{r^{2}sin^{2}\theta} . 
\end{equation} 
We see that components of metric have a different dimension. Similar to the cylindrical coordinate, we use the arbitrary, finite constant length $l$ in $\psi=l\theta$ and $\varphi =l\phi$. So, the iso-dimensional spherical coordinate system is
\begin{equation}\label{20} 
ds^{2}=-c^{2}d\tau^{2}=-d\chi^{2}+dr^{2}+R^{2}d\psi^{2}+R^{2}sin^{2}(\frac{\psi}{l})d^{2}\varphi. 
\end{equation} 
The dimensionless components of $g_{\mu\nu}$ and $g^{\mu\nu}$ are 
\begin{eqnarray}\label{21} 
&&g_{00}=-1,\qquad g_{rr}=1,\qquad g_{\psi\psi}=R^{2},\qquad g_{\varphi\varphi}=R^{2}sin^{2}\theta, 
\nonumber\\ &&g^{00}=-1,\qquad g^{rr}=1,\qquad g^{\psi\psi}=\frac{1}{R^{2}},\qquad g^{\varphi\varphi}=\frac{1}{R^{2}sin^{2}\theta}.\nonumber\\ 
\end{eqnarray} 
In the cylindrical and spherical coordinate systems $l$ may be a free constant which will be chose adopted to each study. For example in a relativistic fluids around the black hole, $r_{g}$ (Schwarzschild radius) may be a suitable chose for $l$, etc. Also, $l$ may be selected as unite ($l=1$), etc. 
\section{Four velocity}\label{s5} 
The components of four velocity and covariant components of four velocity are derived by 
\begin{equation}\label{22} 
u^{\mu}=\frac{dx^{\nu}}{d\tau}=\gamma\frac{dx^{\nu}}{dt}, \qquad u_{\mu}=g_{\mu\nu}u^{\nu}, 
\end{equation} 
where, $\gamma=\frac{dt}{d\tau}=\frac{1}{\sqrt{1-\frac{v^{2}}{c^{2}}}}$ is the Lorentz factor.
In the Lorentz factor, the expansions of $\gamma$, and $\gamma^{2}$ are given by 
\begin{eqnarray}\label{23} 
&&\gamma=\frac{1}{(1-\frac{v^{2}}{c^{2}})^{\frac{1}{2}}}\cong 1+\frac{v^{2}}{2c^{2}}+O(\frac{1}{c^{4}}), 
\nonumber\\ &&\gamma^{2}=\frac{1}{1-\frac{v^{2}}{c^{2}}} \cong 1+\frac{v^{2}}{c^{2}}+O(\frac{1}{c^{4}}). 
\end{eqnarray} 
\subsection{Four velocity in the cylindrical coordinate system} 
The components and covariant components of four velocity in the iso-dimensional cylindrical coordinate system are 
\begin{eqnarray}\label{24} 
&& u^{0}=c\gamma,\qquad u^{r}=\dot{r}\gamma,\qquad u^{\psi}=\dot{\psi}\gamma=l\dot{\theta}\gamma,\qquad u^{z}=\dot{z}\gamma, 
\nonumber\\ &&u_{0}=-c\gamma,\qquad u_{r}=\dot{r}\gamma,\qquad u_{\psi}=R^2l\dot{\theta}\gamma=\frac{ r^2 \dot{\theta}\gamma}{l},\nonumber\\ && u_{z}=\dot{z}\gamma. 
\end{eqnarray} 
\subsection{Four velocity in the spherical coordinate system} 
In the iso-dimensional spherical coordinate system, the components of four velocity and covariant components of four velocity are 
\begin{eqnarray}\label{25} 
&& u^{0}=c\gamma,\qquad u^{r}=\dot{r}\gamma,\qquad u^{\psi}=\dot{\psi}\gamma=l\dot{\theta}\gamma,\qquad u^{\varphi}=l\dot{\phi}\gamma, 
\nonumber\\ &&u_{0}=-c\gamma,\qquad u_{r}=\dot{r}\gamma,\qquad u_{\psi}=R^2l\dot{\theta}\gamma=\frac{ r^2 \dot{\theta}\gamma}{l},\nonumber\\ && u_{\varphi}=\frac{ r^2sin^{2}\theta \dot{\phi}\gamma }{l}. 
\end{eqnarray} 
\section{ Perfect fluid energy momentum tensor} \label{s6} 
In this section the components of perfect fluids energy momentum tensor are calculated in two coordinate systems. 
\subsection{Cylindrical coordinate system} 
The components of perfect fluid energy momentum tensor are calculated in IDS and in the cylindrical coordinate system as 
\begin{eqnarray}\label{26} 
&&T^{00}_{Pf}=\gamma^2\rho c^2 -p,\quad T^{0r}_{Pf}=\gamma^2\rho c\dot{r},\quad T^{0\psi}_{Pf}=l\gamma^2\rho c\dot{\theta}, 
\nonumber\\ && T^{0z}_{Pf}=\gamma^2\rho c\dot{z},\quad 
T^{rr}_{Pf}=\gamma^2\rho \dot{r}^2+p, \quad T^{r\psi}_{Pf}=l\gamma^2\rho \dot{r}\dot{\theta}, 
\nonumber\\ && T^{rz}_{Pf}=\gamma^2\rho \dot{r}\dot{z}, 
\quad T^{\psi\psi}_{Pf}=l^2(\gamma^2\rho \dot{\theta}^2+\frac{p}{r^2}), 
\nonumber\\ && T^{\psi z}_{Pf}=l\gamma^2\rho \dot{\theta}\dot{z},\quad 
T^{zz}_{Pf}=\gamma^2\rho \dot{z}^2+p. 
\end{eqnarray} 
\subsection{Spherical coordinate system} 
In the spherical coordinate system, the components of perfect fluids energy momentum tensor are 
\begin{eqnarray}\label{27} 
&&T^{00}_{Pf}=\gamma^2\rho c^2 -p,\quad T^{0r}_{Pf}=\gamma^2\rho c\dot{r},\quad T^{0\psi}_{Pf}=l\gamma^2\rho c\dot{\theta}, 
\nonumber\\ && T^{0\varphi}_{Pf}=l\gamma^2\rho c\dot{\phi},\quad 
T^{rr}_{Pf}=\gamma^2\rho \dot{r}^2+p, \quad T^{r\psi}_{Pf}=l\gamma^2\rho \dot{r}\dot{\theta}, 
\nonumber\\ && T^{r\varphi}_{Pf}=l\gamma^2\rho \dot{r}\dot{\phi}, 
\quad T^{\psi\psi}_{Pf}=l^2(\gamma^2\rho \dot{\theta}^2+\frac{p}{r^2}), 
\nonumber\\ && T^{\psi \varphi}_{Pf}=l^2\gamma^2\rho \dot{\theta}\dot{\phi},\quad 
T^{\varphi\varphi}_{Pf}=l^2(\gamma^2\rho \dot{\phi}^2+\frac{p}{r^2\sin^2\theta}). 
\end{eqnarray} 
\section{ Shear and bulk tensors}\label{s7} 
\subsection{Shear and bulk tensors in the cylindrical coordinate system} 
In the iso-dimensional cylindrical coordinate system, the non-zero components of the Christoffel symbols are derived as 
\begin{equation}\label{28} 
\Gamma^{r}_{\psi\psi}=-\frac{r}{l^2},\qquad \Gamma^{\psi}_{r\psi}=\frac{1}{r}. 
\end{equation} 
The expansion of the fluid world line in the cylindrical coordinate system is calculated by equation (\ref{8}) as 
\begin{eqnarray}\label{29} 
&&\Theta=\partial_{\gamma}u^{\gamma} +\Gamma_{\gamma\nu}^{\nu}u^{\gamma}=\partial_{\gamma}u^{\gamma}+\frac{\gamma \dot{r}}{r}=\nabla\cdot(\gamma\vec{v} )+\frac{\partial \gamma}{\partial t},\nonumber\\ 
&& \Rightarrow\Theta=\nabla\cdot\vec{v}+\frac{1}{2c^{2}}(\nabla\cdot(v^{2}\vec{v})+\frac{\partial v^{2}}{\partial t})+O(c^{-4}),
\end{eqnarray} 
in the cylindrical coordinate $\nabla\cdot\vec{v}$ is 
\begin{eqnarray}\label{30} 
&&\nabla\cdot\vec{v}=\frac{1}{r}\frac{\partial( r v_{r})}{\partial r}+\frac{1}{r}\frac{\partial v_{\theta}}{\partial\theta}+\frac{\partial v_{z}}{\partial z}=\frac{1}{r}\frac{\partial( r\dot{r})}{\partial r}+\frac{\partial \dot{\theta}}{\partial\theta}+\frac{\partial\dot{z}}{\partial z}, 
\nonumber\\&&\qquad=\frac{1}{r}\frac{\partial( r\dot{r})}{\partial r}+\frac{\partial \dot{\psi}}{\partial\psi}+\frac{\partial\dot{z}}{\partial z}, 
\end{eqnarray} 
where, $ v_{r},v_{\theta},v_{z} $ are the components of velocity. The components of projection tensor are given in the appendix A. Also, the components of the bulk tensor in this coordinate system are calculated as 
\begin{eqnarray}\label{31} 
&&b^{00}=\frac{v^{2}\nabla\cdot\vec{v}}{c^{2}}+O(\frac{1}{c^{4}}),\quad 
\nonumber\\ && b^{0r}=\frac{\dot{r}}{c}(\nabla\cdot\vec{v}+\frac{1}{2c^{2}}[\nabla\cdot(v^{2}\vec{v})+\frac{\partial v^{2}}{\partial t}])+O(\frac{1}{c^{5}}), 
\nonumber\\ && b^{0\psi}=\frac{l\dot{\theta}}{c}(\nabla\cdot\vec{v}+\frac{1}{2c^{2}}[\nabla\cdot(v^{2}\vec{v})+\frac{\partial v^{2}}{\partial t}])+O(\frac{1}{c^{5}}),
\nonumber\\ && b^{0z}=\frac{\dot{z}}{c}(\nabla\cdot\vec{v}+\frac{1}{2c^{2}}[\nabla\cdot(v^{2}\vec{v})+\frac{\partial v^{2}}{\partial t}])+O(\frac{1}{c^{5}}),
\nonumber\\ && 
b^{rr}=(1+\frac{\dot{r}^{2}}{c^{2}})\nabla\cdot\vec{v}+\frac{1}{2c^{2}}[\nabla\cdot(v^{2}\vec{v})+\frac{\partial v^{2}}{\partial t}]+O(\frac{1}{c^{4}}),\nonumber\\ && 
b^{r\psi}=\frac{l\dot{r}\dot{\theta}\nabla\cdot\vec{v}}{c^{2}}+O(\frac{1}{c^{4}}), \quad b^{rz}=\frac{\dot{r}\dot{z}\nabla\cdot\vec{v}}{c^{2}}+O(\frac{1}{c^{4}}),
\nonumber\\ &&b^{\psi\psi}=l^{2}(\frac{1}{r^{2}}+\frac{\dot{\theta}^{2}}{c^{2}})\nabla\cdot\vec{v}+\frac{l^{2}}{2r^{2}c^{2}}[\nabla\cdot(v^{2}\vec{v})+\frac{\partial v^{2}}{\partial t}]+O(\frac{1}{c^{4}}),
\nonumber\\ && b^{\psi z}=\frac{l\dot{\theta}\dot{z}\nabla\cdot\vec{v}}{c^{2}}+O(\frac{1}{c^{4}}),
\nonumber\\ && b^{zz}=(1+\frac{\dot{z}^{2}}{c^{2}})\nabla\cdot\vec{v}+\frac{1}{2c^{2}}[\nabla\cdot(v^{2}\vec{v})+\frac{\partial v^{2}}{\partial t}]+O(\frac{1}{c^{4}}). 
\end{eqnarray} 
The components of four acceleration in this coordinate are obtained as 
\begin{eqnarray}\label{32} 
&&a^{0}=u^{0}_{;\gamma}u^{\gamma}=c\gamma(\nabla\gamma\cdot\vec{v}+\gamma\frac{\partial\gamma}{\partial t}),
\nonumber\\ &&a^{r}=u^{r}_{;\gamma}u^{\gamma}=\gamma\nabla(\gamma \dot{r})\cdot\vec{v}-r\gamma^{2}\dot{\theta}^{2}+\gamma\frac{\partial(\gamma\dot{r})}{\partial t}, 
\nonumber\\ &&a^{\psi}=u^{\psi}_{;\gamma}u^{\gamma}=\gamma l(\nabla(\gamma \dot{\theta} )\cdot\vec{v}+\frac{2\dot{r}\dot{\theta}}{r}+\frac{\partial(\gamma \dot{\theta})}{\partial t}),
\nonumber\\ &&a^{z}=u^{z}_{;\gamma}u^{\gamma}=\gamma\nabla(\gamma\dot{z} )\cdot\vec{v}+\gamma\frac{\partial(\gamma\dot{z})}{\partial t}. 
\end{eqnarray}  
After some calculation the components of shear tensor are derived as 
\begin{eqnarray}\label{33} 
&&\sigma^{00}=\frac{\nabla v^{2}\cdot\vec{v} }{2c^{2}}-\frac{1}{3}b^{00}+O(\frac{1}{c^{4}}) 
\nonumber\\ && \sigma^{0r}=\frac{1}{2c}(\frac{1}{2}\frac{\partial v^{2}}{\partial r}+ \nabla\dot{r}\cdot\vec{v}-r\dot{\theta}^{2}+\frac{\dot{r}\nabla v^{2}\cdot\vec{v}}{2c^{2}}) -\frac{1}{3}b^{0r}+O(\frac{1}{c^{5}}), 
\nonumber\\ && \sigma^{0\psi}=\frac{l}{2c}[\frac{1}{2r^2}\frac{\partial v^{2}}{\partial \theta}+\nabla \dot{\theta}\cdot\vec{v}+\frac{2\dot{r}\dot{\theta}}{r}+\frac{\dot{\theta}\nabla v^{2}\cdot\vec{v}}{2c^{2}}] -\frac{1}{3}b^{0\psi}
+O(\frac{1}{c^{5}}),
\nonumber\\ && \sigma^{0z}=\frac{1}{2c}(\frac{1}{2}\frac{\partial v^{2}}{\partial z}+ \nabla\dot{z}\cdot\vec{v}+\frac{\dot{z}\nabla v^{2}\cdot\vec{v}}{2c^{2}}) -\frac{1}{3}b^{0z}+O(\frac{1}{c^{5}}),
\nonumber\\ && \sigma^{rr}=\frac{\partial (\gamma \dot{r})}{\partial r}+\frac{\dot{r}}{c^{2}}(\nabla \dot{r}\cdot\vec{v}-r\dot{\theta}^{2}+\frac{\partial \dot{r}}{\partial t})-\frac{1}{3}b_{rr}+O(\frac{1}{c^{4}}), 
\nonumber\\ && \sigma^{r\psi}=\frac{l}{2}[\frac{1}{r^2}\frac{\partial (\gamma \dot{r})}{\partial \theta}+\frac{\partial (\gamma \dot{\theta})}{\partial r}+\frac{1}{c^{2}}(\nabla (\dot{\theta}\dot{r})\cdot\vec{v}-r\dot{\theta}^{3}+\frac{2\dot{\theta}\dot{r}^2}{r} \nonumber\\ && \qquad +\frac{\partial( \dot{\theta}\dot{r})}{\partial t})]-\frac{1}{3}b^{r\psi}+O(\frac{1}{c^{4}})
\nonumber\\ && \sigma^{rz}=\frac{1}{2}[\frac{\partial (\gamma \dot{r})}{\partial z}+\frac{\partial (\gamma \dot{z})}{\partial r}+\frac{1}{c^{2}}(\nabla(\dot{r}\dot{z})\cdot\vec{v}-r\dot{\theta}^{2}\dot{z}+\frac{\partial(\dot{z}\dot{r})}{\partial t})] 
  -\frac{1}{3}b^{rz}+O(\frac{1}{c^{4}}), 
\nonumber\\ && \sigma^{\psi\psi}=l^{2}[\frac{1}{r^2}\frac{\partial (\gamma \dot{\theta})}{\partial \theta}+\frac{\gamma\dot{r}}{r}+\frac{\dot{\theta}}{c^{2}}(\nabla \dot{\theta}\cdot\vec{v}+\frac{2\dot{r}\dot{\theta}}{r}+\frac{\partial \dot{\theta}}{\partial t}) -\frac{1}{3}b^{\psi\psi}+O(\frac{1}{c^{4}})
\nonumber\\ && \sigma^{\psi z}=\frac{l}{2}[\frac{\partial (\gamma \dot{\theta})}{\partial z}+\frac{1}{r^{2}}\frac{\partial (\gamma \dot{z})}{\partial \theta}+\frac{1}{c^{2}}(\nabla(\dot{\theta}\dot{z})\cdot\vec{v}+\frac{2\dot{r}\dot{z}\dot{\theta}}{r}+\frac{\partial ( \dot{\theta}\dot{z})}{\partial t})] 
  -\frac{1}{3}b^{\psi z}+O(\frac{1}{c^{4}})
\nonumber\\ && \sigma^{zz}=\frac{\partial (\gamma \dot{z})}{\partial z}+\frac{\dot{z}}{c^{2}}(\nabla \dot{z}\cdot\vec{v}+\frac{\partial\dot{z}}{\partial t})-\frac{1}{3}b^{zz}+O(\frac{1}{c^{4}}), 
\end{eqnarray} 
\subsection{Shear and bulk tensors in the spherical coordinate} 
the non-zero components of the Christoffel symbols in the iso-dimension spherical coordinate system are 
\begin{eqnarray}\label{34} 
&&\Gamma^{r}_{\psi\psi}=-\frac{r}{l^2},\quad\Gamma^{r}_{\varphi\varphi}=-\frac{r\sin^2{\theta}}{l^2},\quad\Gamma^{\psi}_{\varphi\varphi}=-\frac{\sin\theta \cos\theta}{l}, 
\nonumber\\ 
&& \Gamma^{\psi}_{r\psi}= \frac{1}{r}, \qquad \Gamma^{\varphi}_{r\varphi}=\frac{1}{r},\qquad \Gamma^{\psi}_{\psi\varphi}=\frac{\cos\theta}{l\sin\theta}. 
\end{eqnarray} 
We use equation (\ref{8}) to calculate the expansion of the fluid world line in the spherical coordinate system 
\begin{eqnarray}\label{35} 
&&\Theta=\partial_{\gamma}u^{\gamma} +\Gamma_{\gamma\nu}^{\nu}u^{\gamma}=\partial_{\gamma}u^{\gamma}+\frac{2\gamma \dot{r}}{r}+\gamma\dot{\theta}\cot\theta 
\nonumber\\&&\quad =\nabla\cdot(\gamma\vec{v} ) +\frac{\partial \gamma}{\partial t} 
,\nonumber\\ 
&& \Rightarrow\Theta=\nabla\cdot\vec{v}+\frac{1}{2c^{2}}(\nabla\cdot(v^{2}\vec{v})+\frac{\partial v^{2}}{\partial t})+O(\frac{1}{c^{4}}). 
\end{eqnarray} 
where, in the spherical coordinate 
$\nabla\cdot\vec{v}=\frac{1}{r^2}\frac{\partial (r^2 v_{r})}{\partial r}+\frac{1}{r\sin\theta}\frac{\partial( \sin\theta v_{\theta})}{\partial\theta}+\frac{1}{r\sin\theta}\frac{\partial v_{\phi}}{\partial \phi}$. 
Projection tensor are given in the appendix A; so, the components of the bulk tensor in the iso-dimension spherical coordinate system are calculated as \vspace{-1cm} 
\begin{eqnarray}\label{36} 
&&b^{00}=\frac{v^{2}\nabla\cdot\vec{v}}{c^{2}}+O(\frac{1}{c^{4}}),
\nonumber\\ && b^{r0}=\frac{\dot{r}}{c}(\nabla\cdot\vec{v}+\frac{1}{2c^{2}}[\nabla\cdot(v^{2}\vec{v})+\frac{\partial v^{2}}{\partial t}])+O(\frac{1}{c^{5}}), 
\nonumber\\ &&  
b^{0\psi}=\frac{l\dot{\theta}}{c}(\nabla\cdot\vec{v}+\frac{1}{2c^{2}}[\nabla\cdot(v^{2}\vec{v})+\frac{\partial v^{2}}{\partial t}])+O(\frac{1}{c^{5}}),
\nonumber\\ && b^{0\varphi}=\frac{l\dot{\phi}}{c}(\nabla\cdot\vec{v}+\frac{1}{2c^{2}}[\nabla\cdot(v^{2}\vec{v})+\frac{\partial v^{2}}{\partial t}])+O(\frac{1}{c^{5}}), 
\nonumber\\ &&b^{rr}=(1+\frac{\dot{r}^{2}}{c^{2}})\nabla\cdot\vec{v}+\frac{1}{2c^{2}}[\nabla\cdot(v^{2}\vec{v})+\frac{\partial v^2}{\partial t}]+O(\frac{1}{c^{4}}), 
\nonumber\\ && b^{r\psi}=\frac{l\dot{r}\dot{\theta}\nabla\cdot\vec{v}}{c^{2}}+O(\frac{1}{c^{4}}),\quad 
b^{r\varphi}=\frac{l\dot{r}\dot{\phi}\nabla\cdot\vec{v}}{c^{2}}+O(\frac{1}{c^{4}}), 
\nonumber\\ &&b^{\psi\psi}=l^{2}(\frac{1}{r^{2}}+\frac{\dot{\theta}^{2}}{c^{2}})\nabla\cdot\vec{v}+\frac{l^{2}}{2c^2r^{2}}[\nabla\cdot(v^{2}\vec{v})+\frac{\partial v^2}{\partial t}]+O(\frac{1}{c^{4}}),\qquad 
\nonumber\\ &&b^{\psi\varphi }=\frac{l^2\dot{\theta}\dot{\phi}\nabla\cdot\vec{v}}{c^{2}}+O(\frac{1}{c^{4}}), 
\nonumber\\ &&b^{\varphi\varphi}=l^{2}(\frac{1}{r^{2}sin^{2}\theta}+\frac{\dot{\phi}^{2}}{c^{2}})\nabla\cdot\vec{v}+\frac{l^{2}}{2 c^{2}r^{2}sin^{2}\theta} 
[\nabla\cdot(v^{2}\vec{v})+\frac{\partial v^2}{\partial t}]
+O(\frac{1}{c^{4}}). 
\end{eqnarray}  
The components of four acceleration are derived with equation (\ref{8}) as \vspace{-2cm} 
\begin{eqnarray}\label{37} 
&&a^{0}=c\gamma(\nabla\gamma\cdot\vec{v}+\gamma\frac{\partial\gamma}{\partial t}),
\nonumber\\ &&a^{r}=\gamma\nabla(\gamma \dot{r})\cdot\vec{v}-r\gamma^{2}\dot{\theta}^{2}-r\gamma^{2}\sin^{2}\theta\dot{\phi}^{2}+\gamma\frac{\partial(\gamma\dot{r})}{\partial t}, 
\nonumber\\ &&a^{\psi}=\gamma l(\nabla(\gamma \dot{\theta} )\cdot\vec{v}+\frac{2\gamma\dot{r}\dot{\theta}}{r}-\gamma\sin\theta \cos\theta\dot{\phi}^{2}+\frac{\partial(\gamma \dot{\theta})}{\partial t}), 
\nonumber\\ &&a^{\varphi }=\gamma l(\gamma\nabla(\gamma \dot{\phi})\cdot\vec{v}+\frac{2\gamma\dot{r}\dot{\phi}}{r}+2\gamma\cot\theta \dot{\phi}\dot{\theta} 
+\frac{\partial(\gamma \dot{\phi})}{\partial t}).
\end{eqnarray} 
The components of shear tensor are calculated as 
\begin{eqnarray}\label{38} 
&&\sigma^{00}=\frac{\nabla v^{2}\cdot\vec{v} }{2c^{2}}-\frac{1}{3}b^{00}+O(\frac{1}{c^{4}}) 
\nonumber\\ && \sigma^{0r}=\frac{1}{2c}[\frac{1}{2}\frac{\partial v^{2}}{\partial r}+ \nabla\dot{r}\cdot\vec{v}-r\dot{\theta}^{2}-r\sin^{2}\theta\dot{\phi}^{2}+\frac{\dot{r}\nabla v^{2}\cdot\vec{v}}{2c^{2}}] -\frac{1}{3}b^{0r}
+O(\frac{1}{c^{5}}),
\nonumber\\ && \sigma^{0\psi}=\frac{l}{2c}[\frac{1}{2r^2}\frac{\partial v^{2}}{\partial \theta}+\nabla \dot{\theta}\cdot\vec{v}+\frac{2\dot{r}\dot{\theta}}{r}-\sin\theta\cos\theta\dot{\phi}^{2}+\frac{\dot{\theta}\nabla v^{2}\cdot\vec{v}}{2c^{2}}] 
-\frac{1}{3}b^{0\psi}+O(\frac{1}{c^{5}}), 
\nonumber\\ && \sigma^{0\varphi}=\frac{l}{2c}[\frac{1}{2r^2\sin^{2}\theta}\frac{\partial v^{2}}{\partial \phi} 
+\nabla \dot{\phi}\cdot\vec{v}+\frac{2\dot{r}\dot{\phi}}{r}+2\cot\theta\dot{\theta}\dot{\phi}+\frac{\dot{\phi}\nabla v^{2}\cdot\vec{v}}{2c^{2}}] 
 -\frac{1}{3}b^{0\varphi}+O(\frac{1}{c^{5}}),
\nonumber\\ && \sigma^{rr}=\frac{\partial (\gamma \dot{r})}{\partial r}+\frac{\dot{r}}{c^{2}}(\nabla \dot{r}\cdot\vec{v}-r\dot{\theta}^{2}-r\sin^{2}\theta\dot{\phi}^{2}+\frac{\partial \dot{r}}{\partial t}) -\frac{1}{3}b^{rr}+O(\frac{1}{c^{4}}),
\nonumber\\ && \sigma^{r\psi}=\frac{l}{2}[\frac{1}{r^2}\frac{\partial (\gamma \dot{r})}{\partial \theta}+\frac{\partial (\gamma \dot{\theta})}{\partial r}+\frac{1}{c^{2}}(\nabla(\dot{\theta} \dot{r})\cdot\vec{v}+\frac{2\dot{\theta} \dot{r}^2}{r} -\sin\theta\cos\theta\dot{r}\dot{\phi}^{2}
 -r\dot{\theta}^{3}-r\sin^2\theta\dot{\theta}\dot{\phi}^{2}+\frac{\partial(\dot{r}\dot{\theta})}{\partial t})] 
\nonumber\\ &&\qquad -\frac{1}{3}b^{r\psi}+O(\frac{1}{c^{4}}), 
\nonumber\\ && \sigma^{r\varphi}=\frac{l}{2}[\frac{1}{r^{2}\sin^{2}\theta}\frac{\partial (\gamma \dot{r})}{\partial \phi}+\frac{\partial (\gamma \dot{\phi})}{\partial r}+\frac{1}{c^{2}}(\nabla(\dot{r}\dot{\phi})\cdot\vec{v}+\frac{2\dot{r}^2\dot{\phi}}{r}
 +2\cot\theta\dot{r}\dot{\theta}\dot{\phi} 
-r\sin^{2}\theta\dot{\phi}^{3}-r\dot{\theta}^{2}\dot{\phi} 
+\frac{\partial(\dot{r}\dot{\phi})}{\partial t}]
\nonumber\\ &&\qquad-\frac{1}{3}b^{r\varphi}+O(\frac{1}{c^{4}}), 
\nonumber\\ && \sigma^{\psi\psi}=l^2[\frac{1}{r^2}\frac{\partial (\gamma \dot{\theta})}{\partial \theta}+\frac{\gamma\dot{r}}{r}+\frac{\dot{\theta}}{c^{2}}(\nabla \dot{\theta}\cdot\vec{v}+\frac{2\dot{r}\dot{\theta}}{r} 
-\sin\theta\cos\theta\dot{\phi}^2+\frac{\partial\dot{\theta}}{\partial t})]
\nonumber\\ &&\qquad-\frac{1}{3}b^{\psi\psi}+O(\frac{1}{c^{4}}), 
\nonumber\\ && \sigma^{\psi\varphi}=\frac{l^2}{2}[\frac{1}{r^2\sin^2\theta}\frac{\partial (\gamma \dot{\theta})}{\partial \phi}+\frac{1}{r^2}\frac{\partial (\gamma \dot{\phi})}{\partial \theta}
+\frac{1}{c^{2}}(\nabla(\dot{\theta}\dot{\phi})\cdot\vec{v} 
+\frac{4\dot{r}\dot{\theta}\dot{\phi}}{r} 
  -\sin\theta\cos\theta\dot{\phi}^3+2\cot\theta\dot{\phi}\dot{\theta}^2+ \frac{\partial(\dot{\theta}\dot{\phi})}{\partial t})] 
\nonumber\\ &&\qquad -\frac{1}{3}b^{\psi\varphi}+O(\frac{1}{c^{4}}), 
\nonumber\\ && \sigma^{\varphi\varphi}=l^2[\frac{1}{r^{2}\sin^{2}\theta}(\frac{\partial (\gamma \dot{\phi})}{\partial \phi}+ \frac{\gamma\dot{r}}{r}+\gamma\cot\theta\dot{\theta}) +\frac{\dot{\phi}}{c^{2}}(\nabla \dot{\phi}\cdot\vec{v}+\frac{2\dot{r}\dot{\phi}}{r} 
+2\cot\theta\dot{\theta}\dot{\phi} 
+\frac{\partial\dot{\phi}}{\partial t}] 
 -\frac{1}{3}b^{\varphi\varphi}+O(\frac{1}{c^{4}}).\nonumber\\
\end{eqnarray} 
\section{Shear stress viscosity energy momentum }\label{s8} 
The energy momentum of shear stress viscosity are derived from equations (3)-(9) as 
\begin{equation}\label{39} 
T^{\mu\nu}_{vis}=-2\lambda\sigma^{\mu\nu}-\zeta b^{\mu\nu}, 
\end{equation} 
in the general cases, the coefficients of the dynamical viscosity and bulk viscosity( $ \lambda $ and $\zeta $) are not constant. These coefficients may be calculated by the thermodynamics variables (for example see \cite{we} ). 
In the Newtonian fluids $\zeta=-\frac{2}{3}\lambda$, therefore we have 
\begin{equation}\label{40} 
T^{\mu\nu}_{vis}=-\lambda [g^{\nu\beta}u^{\mu}_{;\beta}+g^{\mu\alpha}u^{\nu}_{;\alpha}+\frac{1}{c^{2}}(a^{\mu}u^{\nu}+a^{\nu}u^{\mu})]+\frac{4\lambda}{3}b^{\mu\nu}. 
\end{equation} 
The components of shear stress viscosity energy momentum tensor are calculated in two coordinate systems. 
\subsection{In the cylindrical coordinate system} 
The components of shear stress viscosity energy momentum tensor in cylindrical coordinate system are calculated as  
\begin{eqnarray}\label{41} 
&&T^{00}_{vis}=-\lambda\frac{\nabla v^{2}\cdot\vec{v} }{c^{2}}-(\zeta-\frac{2\lambda}{3})\frac{v^{2}\nabla\cdot\vec{v}}{c^{2}}+O(\frac{1}{c^{4}}), 
\nonumber\\ && T^{0r}_{vis}=-\frac{\lambda}{c}(\frac{1}{2}\frac{\partial v^{2}}{\partial r}+\nabla\dot{r}\cdot\vec{v} -r\dot{\theta}^{2}+\frac{\dot{r}\nabla v^{2}\cdot\vec{v}}{2c^{2}})-\frac{\dot{r}}{c}(\zeta-\frac{2\lambda}{3})
 [\nabla\cdot\vec{v}+\frac{1}{2c^{2}}(\nabla\cdot(v^{2}\vec{v})+\frac{\partial v^{2}}{\partial t})]
\nonumber\\ && T^{0\psi}_{vis}=-\frac{l\lambda}{cr^2}[\frac{1}{2}\frac{\partial v^{2}}{\partial \theta}+\nabla (r^{2}\dot{\theta})\cdot\vec{v}+\frac{\dot{\theta}\nabla v^{2}\cdot\vec{v}}{2c^{2}}]-\frac{l\dot{\theta}}{c}(\zeta-\frac{2\lambda}{3})
 [\nabla\cdot\vec{v}+\nabla\cdot\vec{v}+\frac{1}{2c^{2}}(\nabla\cdot(v^{2}\vec{v})+\frac{\partial v^{2}}{\partial t})]\nonumber\\ &&\qquad+O(\frac{1}{c^{5}}), 
\nonumber\\ && T^{0z}_{vis}=-\frac{\lambda}{c}(\frac{1}{2}\frac{\partial v^{2}}{\partial z}- \nabla\dot{z}\cdot\vec{v}+\frac{\dot{z}\nabla v^{2}\cdot\vec{v}}{2c^{2}})-\frac{\dot{z}}{c}(\zeta-\frac{2\lambda}{3})
[\nabla\cdot\vec{v}+\frac{1}{2c^{2}}(\nabla\cdot(v^{2}\vec{v})+\frac{\partial v^{2}}{\partial t})]+O(\frac{1}{c^{5}}), 
\nonumber\\ && T^{rr}_{vis}=-2\lambda[\frac{\partial (\gamma \dot{r})}{\partial r}+\frac{\dot{r}}{c^{2}}(\nabla \dot{r}\cdot\vec{v}-r\dot{\theta}^2+\frac{\partial \dot{r}}{\partial t})] -(\zeta-\frac{2\lambda}{3})
 [(1+\frac{\dot{r}^{2}}{c^{2}})\nabla\cdot\vec{v}+\frac{1}{2c^{2}}(\nabla\cdot(v^{2}\vec{v})+\frac{\partial v^{2}}{\partial t})]\nonumber\\ &&\qquad+O(\frac{1}{c^{4}}), %
\nonumber\\ && T^{r\psi}_{vis}=-\lambda l[\frac{1}{r^2}\frac{\partial (\gamma \dot{r})}{\partial \theta}+\frac{\partial (\gamma \dot{\theta})}{\partial r}+\frac{1}{c^{2}}(\nabla (\dot{\theta}\dot{r})\cdot\vec{v}-r\dot{\theta}^{3}+\frac{2\dot{\theta}\dot{r}^2}{r}  
 +\frac{\partial( \dot{\theta}\dot{r})}{\partial t})]-(\zeta-\frac{2\lambda}{3})\frac{l\dot{r}\dot{\theta}\nabla\cdot\vec{v}}{c^2} +O(\frac{1}{c^{4}}),
\nonumber\\ && T^{rz}_{vis}=-\lambda[\frac{\partial (\gamma \dot{r})}{\partial z}+\frac{\partial (\gamma \dot{z})}{\partial r}+\frac{1}{c^{2}}(\nabla (\dot{z}\dot{r})\cdot\vec{v}+\frac{\partial(\dot{z}\dot{r})}{\partial t} -r\dot{\theta}^{2}\dot{z}))]
 -(\zeta-\frac{2\lambda}{3})\frac{\dot{r}\dot{z}\nabla\cdot\vec{v}}{c^{2}} +O(\frac{1}{c^{4}}), 
\nonumber\\ && T^{\psi\psi}_{vis}=-2\lambda l^{2}[\frac{1}{r^2}\frac{\partial (\gamma \dot{\theta})}{\partial \theta}+\frac{\gamma\dot{r}}{r}+\frac{\dot{\theta}}{c^{2}}(\nabla \dot{\theta}\cdot\vec{v}+\frac{2\dot{r}\dot{\theta}}{r} +\frac{\partial \dot{\theta}}{\partial t})] 
 -l^2(\zeta-\frac{2\lambda}{3})
[(\frac{1}{r^{2}}+\frac{\dot{\theta}^{2}}{c^{2}}) 
\nabla\cdot\vec{v} 
 \nonumber\\ &&\qquad +\frac{1}{2r^{2}c^{2}}(\nabla\cdot(v^{2}\vec{v}) 
+\frac{\partial v^{2}}{\partial t})] +O(\frac{1}{c^{4}}),\qquad 
\nonumber\\ && T^{\psi z}_{vis}=-\lambda l[\frac{\partial (\gamma \dot{\theta})}{\partial z}+\frac{1}{r^{2}}\frac{\partial (\gamma \dot{z})}{\partial \theta}+\frac{1}{c^{2}}(\nabla(\dot{\theta}\dot{z})\cdot\vec{v} 
+\frac{2\dot{r}\dot{z}\dot{\theta}}{r} 
+\frac{\partial ( \dot{\theta}\dot{z})}{\partial t})] 
-(\zeta-\frac{2\lambda}{3})\frac{l\dot{\theta}\dot{z}\nabla\cdot\vec{v}}{c^{2}}+O(\frac{1}{c^{4}}), 
\nonumber\\ && T^{zz}_{vis}=-2\lambda[\frac{\partial (\gamma \dot{z})}{\partial z}+\frac{\dot{z}}{c^{2}}(\nabla \dot{z}\cdot\vec{v}+\frac{\partial\dot{z}}{\partial t})]-(\zeta-\frac{2\lambda}{3}) 
 [(1+\frac{\dot{z}^{2}}{c^{2}})\nabla\cdot\vec{v}+\frac{1}{2c^{2}}(\nabla\cdot(v^{2}\vec{v})+\frac{\partial v^{2}}{\partial t})]+O(\frac{1}{c^{4}}). \nonumber\\ 
\end{eqnarray} 
\subsection{In the spherical coordinate system} 
In the spherical coordinate the components of shear stress viscosity energy momentum tensor are derived as 
\begin{eqnarray}\label{42} 
&& T^{00}_{vis}=-\lambda(\frac{\nabla v^{2}\cdot\vec{v} }{c^{2}})-(\zeta-\frac{2\lambda}{3})\frac{v^{2}\nabla\cdot\vec{v}}{c^{2}}+O(\frac{1}{c^{4}}), 
\nonumber\\ && T^{0r}_{vis}=\frac{\lambda}{c}(\frac{1}{2}\frac{\partial v^{2}}{\partial r}-\nabla\dot{r}\cdot\vec{v} +r\dot{\theta}^{2}+r\sin^{2}\theta\dot{\phi}^{2}+\frac{\dot{r}\nabla v^{2}\cdot\vec{v}}{2c^{2}}) 
-\frac{\dot{r}}{c}(\zeta-\frac{2\lambda}{3})[\nabla\cdot\vec{v}+\nabla\cdot(v^{2}\vec{v})+\frac{\partial v^{2}}{\partial t}]\nonumber\\ &&\qquad +O(\frac{1}{c^{5}}), 
\nonumber\\ && T^{0\psi}_{vis}=-\frac{\lambda l}{c}[\frac{1}{2r^2}\frac{\partial v^{2}}{\partial \theta}+\nabla \dot{\theta}\cdot\vec{v}+\frac{2\dot{r}\dot{\theta}}{r}-\sin\theta\cos\theta\dot{\phi}^{2}] 
-\frac{l\dot{\theta}}{c}(\zeta-\frac{2\lambda}{3})[\nabla\cdot\vec{v}+\nabla\cdot(v^{2}\vec{v})+\frac{\partial v^{2}}{\partial t}]\nonumber\\ &&\qquad +O(\frac{1}{c^{5}}), 
\nonumber\\ && T^{0\varphi}_{vis}=-\frac{\lambda l}{c}[\frac{1}{2r^2\sin^{2}\theta}\frac{\partial v^{2}}{\partial \phi} 
+\nabla \dot{\phi}\cdot\vec{v}+\frac{2\dot{r}\dot{\phi}}{r}+2\cot\theta\dot{\theta}\dot{\phi}] 
 -\frac{l\dot{\phi}}{c}(\zeta-\frac{2\lambda}{3})[\nabla\cdot\vec{v}+\nabla\cdot(v^{2}\vec{v})+\frac{\partial v^{2}}{\partial t}]\nonumber\\ &&\qquad+O(\frac{1}{c^{5}}), 
\nonumber\\ && T^{rr}_{vis}=-2\lambda[\frac{\partial (\gamma \dot{r})}{\partial r}+\frac{\dot{r}}{c^{2}}(\nabla \dot{r}\cdot\vec{v}-r\dot{\theta}^{2}-r\sin^{2}\theta\dot{\phi}^{2}+\frac{\partial \dot{r}}{\partial t})] 
 -(\zeta-\frac{2\lambda}{3})[(1+\frac{\dot{r}^{2}}{c^{2}})\nabla\cdot\vec{v} 
\nonumber\\ &&\qquad +\frac{1}{2c^{2}}(\nabla\cdot(v^{2}\vec{v})+\frac{\partial v^2}{\partial t})]+O(\frac{1}{c^{4}}), 
\nonumber\\ && T^{r\psi}_{vis}=-\lambda l[\frac{1}{r^2}\frac{\partial (\gamma \dot{r})}{\partial \theta}+\frac{\partial (\gamma \dot{\theta})}{\partial r}+\frac{1}{c^{2}}(\nabla(\dot{\theta} \dot{r})\cdot\vec{v}+\frac{2\dot{\theta} \dot{r}^2}{r}-r\dot{\theta}^{3}
 -\sin\theta\cos\theta\dot{r}\dot{\phi}^{2} -r\sin^2\theta\dot{\theta}\dot{\phi}^{2}+\frac{\partial(\dot{r}\dot{\theta})}{\partial t})] 
\nonumber\\ &&\qquad -(\zeta-\frac{2\lambda}{3})\frac{l\dot{r}\dot{\phi}\nabla\cdot\vec{v}}{c^{2}}
+O(\frac{1}{c^{4}}), 
\nonumber\\ && T^{r\varphi}_{vis}=-\lambda l[\frac{1}{r^{2}\sin^{2}\theta}\frac{\partial (\gamma \dot{r})}{\partial \phi}+\frac{\partial (\gamma \dot{\phi})}{\partial r}+\frac{1}{c^{2}}(\nabla(\dot{r}\dot{\phi})\cdot\vec{v}-\cot\theta\dot{r}\dot{\theta}\dot{\phi}
 +\frac{2\dot{r}^2\dot{\phi}}{r} -r\sin^{2}\theta\dot{\phi}^{3}-r\dot{\theta}^{2}\dot{\phi}
\nonumber\\ &&\qquad+\frac{\partial(\dot{r}\dot{\phi})}{\partial t}] 
 -(\zeta-\frac{2\lambda}{3})\frac{l\dot{r}\dot{\phi}\nabla\cdot\vec{v}}{c^{2}}
 +O(\frac{1}{c^{4}}), 
\nonumber\\ && T^{\psi\psi}_{vis}=-2\lambda l^2[\frac{1}{r^2}\frac{\partial (\gamma \dot{\theta})}{\partial \theta}+\frac{\gamma\dot{r}}{r}+\frac{\dot{\theta}}{c^{2}}(\nabla \dot{\theta}\cdot\vec{v}+\frac{2\dot{r}\dot{\theta}}{r} -\sin\theta\cos\theta\dot{\phi}^2 
 +\frac{\partial\dot{\theta}}{\partial t})]
-l^2(\zeta-\frac{2\lambda}{3}) 
[(\frac{1}{r^{2}}+\frac{\dot{\theta}^{2}}{c^{2}})\nabla\cdot\vec{v}
\nonumber\\ &&\qquad +\frac{1}{2c^2r^{2}}(\nabla\cdot(v^{2}\vec{v})+\frac{\partial v^2}{\partial t})]
 +O(\frac{1}{c^{4}}), 
\nonumber\\ && T^{\psi\varphi}_{vis}=-\lambda l^2[\frac{1}{r^2\sin^2\theta}\frac{\partial (\gamma \dot{\theta})}{\partial \phi}+\frac{1}{r^2}\frac{\partial (\gamma \dot{\phi})}{\partial \theta}
+\frac{1}{c^{2}}(\nabla(\dot{\theta}\dot{\phi})\cdot\vec{v} 
+\frac{4\dot{r}\dot{\theta}\dot{\phi}}{r} 
 -\sin\theta\cos\theta\dot{\phi}^3 
 +\cot\theta\dot{\phi}\dot{\theta}^2+ \frac{\partial(\dot{\theta}\dot{\phi})}{\partial t})] 
\nonumber\\ &&\qquad-(\zeta-\frac{2\lambda}{3})\frac{l^2\dot{\theta}\dot{\phi}\nabla\cdot\vec{v}}{c^{2}}
 +O(\frac{1}{c^{4}}), 
\nonumber\\ && T^{\varphi\varphi}_{vis}=-2\lambda l^2[\frac{1}{r^{2}\sin^{2}\theta}(\frac{\partial (\gamma \dot{\phi})}{\partial \phi}+ \frac{\gamma\dot{r}}{r}+\gamma\cot\theta\dot{\theta}) +\frac{\dot{\phi}}{c^{2}}(\nabla \dot{\phi}\cdot\vec{v}+\frac{2\dot{r}\dot{\phi}}{r} 
+2\cot\theta\dot{\theta}\dot{\phi} 
+\frac{\partial\dot{\phi}}{\partial t}] 
\nonumber\\ &&\qquad -l^2(\zeta-\frac{2\lambda}{3}) 
[(\frac{1}{r^{2}sin^{2}\theta}+\frac{\dot{\phi}^{2}}{c^{2}})\nabla\cdot\vec{v}
+\frac{1}{2 c^{2}r^{2}sin^{2}\theta}(\nabla\cdot(v^{2}\vec{v}) +\frac{\partial v^2}{\partial t})]+O(\frac{1}{c^{4}}).
\end{eqnarray} 
\section{Heat flux of the relativistic fluids}\label{s9} 
The heat flux is an important factor to energy transfer in the relativistic fluid. So, in this section the components of heat flux vector of a relativistic fluid are derived. We use the equation (\ref{11}) to calculate the heat flux. After some calculation the components of the heat flux in the relativistic state and in two coordinate systems are derived. 
\subsection{Heat flux in the cylindrical coordinate system} 
The components of the heat flux of a relativistic fluid in the cylindrical coordinate system are calculated as 
\begin{eqnarray}\label{43} 
&&q^{0}=- \frac{\kappa}{c}(\frac{v^2}{c^2}\frac{\partial T}{\partial t}+\nabla T\cdot \vec{v}+\frac{T}{c^2}( a_{r}\dot{r}+\dot{\theta}a_{\theta}+\dot{z}a_{z}))+O(\frac{1}{c^{5}}), 
\nonumber\\ && q^{r}=-\kappa [\frac{\partial T}{\partial r}+\frac{T a_{r}}{c^{2}}
+\frac{\dot{r}}{c^{2}}(\nabla T\cdot \vec{v}+\frac{\partial T}{\partial t})]+O(\frac{1}{c^{4}}), 
\nonumber\\ && q^{\psi}=-l\kappa [\frac{1}{r^{2}}(\frac{\partial T}{\partial \theta}+\frac{T a_{\theta}}{ c^{2}})+\frac{\dot{\theta}}{c^2}(\nabla T\cdot \vec{v} 
+\frac{\partial T}{\partial t})]+O(\frac{1}{c^{4}}), 
\nonumber\\ && q^{z}=-\kappa [\frac{\partial T}{\partial z}+\frac{T a_{z}}{c^{2}}+\frac{\dot{z}}{c^{2}}(\nabla T\cdot \vec{v}+\frac{\partial T}{\partial t})]+O(\frac{1}{c^{4}}).
\end{eqnarray} 
\subsection{Heat flux in the spherical coordinate system} 
The components of heat flux in the spherical coordinate system are given as 
\begin{eqnarray}\label{44} 
&&q^{0}=- \frac{\kappa}{c}(\frac{v^2}{c^2}\frac{\partial T}{\partial t}+\nabla T\cdot \vec{v}+\frac{T}{c^2}(\dot{r}a_{r}+\dot{\theta}a_{\theta}+\dot{z}a_{z}))+O(\frac{1}{c^{5}}), 
\nonumber\\ && q^{r}=-\kappa [\frac{\partial T}{\partial r}+\frac{Ta_{r}}{c^{2}}
+\frac{\dot{r}}{c^{2}}(\nabla T\cdot \vec{v}+\frac{\partial T}{\partial t})]+O(\frac{1}{c^{4}}), 
\nonumber\\ && q^{\psi}=-l\kappa [\frac{1}{r^{2}}(\frac{\partial T}{\partial \theta}+\frac{T a_{\theta}}{c^{2}})+\frac{\dot{\theta}}{c^2}(\nabla T\cdot \vec{v} 
+\frac{\partial T}{\partial t})]+O(\frac{1}{c^{4}}), 
\nonumber\\ && q^{\varphi}=-l\kappa [\frac{1}{r^{2}\sin^{2}\theta}(\frac{\partial T}{\partial \phi}+\frac{T a_{\phi}}{ c^{2}})+\frac{\dot{\phi}}{c^2}(\nabla T\cdot \vec{v}+\frac{\partial T}{\partial t})]+O(\frac{1}{c^{4}}). 
\end{eqnarray}  
\section{Heat flux energy momentum tensor }\label{s10} 
In this section the components of heat flux energy momentum tensor in two coordinate systems are derived.
\subsection{Heat flux energy momentum tensor in the cylindrical coordinate} 
From equations (\ref{10}),(\ref{24}) and (\ref{43}), the components of heat flux energy momentum tensor in cylindrical coordinate are obtained as 
\begin{eqnarray}\label{45} 
&&T^{00}_{heat}=- \frac{2\kappa\gamma}{c^2}\nabla T\cdot \vec{v}+o(\frac{1}{c^{4}}), 
\nonumber\\ && T^{0r}_{heat} 
=-\frac{\kappa}{c}[\gamma\frac{\partial T}{\partial r}+\frac{T a_{r}}{c^{2}}
+\frac{\dot{r}}{c^{2}}(2\nabla T\cdot \vec{v}+\frac{\partial T}{\partial t})]+O(\frac{1}{c^{5}}),
\nonumber\\ && T^{0\psi}_{heat}
=-\frac{\kappa l}{c}[\frac{ 1}{r^{2}}(\gamma\frac{\partial T}{\partial \theta}+\frac{T a_{\theta}}{c^{2}})+\frac{\dot{\theta}}{c^2}(2\nabla T\cdot \vec{v} 
+\frac{\partial T}{\partial t})]+O(\frac{1}{c^{5}}), 
\nonumber\\ && T^{0z}_{heat}
=-\frac{\kappa}{c}[\gamma\frac{\partial T}{\partial z}+\frac{T a_{z}}{c^{2}}+\frac{\dot{z}}{c^{2}}(2\nabla T\cdot \vec{v}+\frac{\partial T}{\partial t})]+O(\frac{1}{c^{5}}), 
\nonumber\\ && T^{rr}_{heat}
=-2\frac{\kappa\dot{r}}{c^{2}}\frac{\partial T}{\partial r}+O(\frac{1}{c^{4}}), 
\nonumber\\ && T^{r\psi}_{heat}
=-\frac{\kappa l}{c^{2}} (\dot{\theta}\frac{\partial T}{\partial r}+\frac{\dot{r}}{r^2}\frac{\partial T}{\partial \theta})+O(\frac{1}{c^{4}}), 
\nonumber\\ && T^{rz}_{heat}
=-\frac{\kappa }{c^{2}}(\dot{z}\frac{\partial T}{\partial r}+\dot{r}\frac{\partial T}{\partial z})+O(\frac{1}{c^{4}}), 
\nonumber\\ && T^{\psi\psi}_{heat}
=-2\frac{\kappa\dot{\theta}l^2 }{r^2c^{2}} \frac{\partial T}{\partial \theta}+O(\frac{1}{c^{4}}), 
\nonumber\\ && T^{\psi z}_{heat}
=-\frac{\kappa l }{c^{2}}(\frac{\dot{z}}{r^{2}}\frac{\partial T}{\partial \theta}+\dot{\theta}\frac{\partial T}{\partial z})+O(\frac{1}{c^{4}}), 
\nonumber\\ && T^{zz}_{heat}
=-2\frac{\kappa\gamma\dot{z} }{c^{2}}\frac{\partial T}{\partial z}+O(\frac{1}{c^{4}}). 
\end{eqnarray} 
\subsection{Heat flux energy momentum tensor in the spherical coordinate} 
From equations (\ref{10}), (\ref{25}) and (\ref{44}), the components of heat flux energy momentum tensor in spherical coordinate are calculated as 
\begin{eqnarray}\label{46} 
&&T^{00}_{heat}
=- \frac{2\kappa\gamma}{c^2}\nabla T\cdot \vec{v}+o(\frac{1}{c^{4}}),
\nonumber\\ && T^{0r}_{heat}
=-\frac{\kappa l}{c}[(\gamma\frac{\partial T}{\partial r}+\frac{T a_{r}}{c^{2}})+\frac{\dot{r}}{c^2}(2\nabla T\cdot \vec{v} 
+\frac{\partial T}{\partial t})]+O(\frac{1}{c^{5}}), 
\nonumber\\ && T^{0\psi}_{heat}
=-\frac{\kappa l}{c}[\frac{ 1}{r^{2}}(\gamma\frac{\partial T}{\partial \theta}+\frac{T a_{\theta}}{c^{2}})+\frac{\dot{\theta}}{c^2}(2\nabla T\cdot \vec{v} 
+\frac{\partial T}{\partial t})]+O(\frac{1}{c^{5}}), 
\nonumber\\ && T^{0\varphi}_{heat}
=-\frac{\kappa l}{c}[\frac{ 1}{r^{2}\sin^{2}\theta}(\gamma\frac{\partial T}{\partial \phi}+\frac{T a_{\phi}}{c^{2}})+\frac{\dot{\phi}}{c^2}(2\nabla T\cdot \vec{v} 
+\frac{\partial T}{\partial t})]+O(\frac{1}{c^{5}}), 
\nonumber\\ && T^{rr}_{heat}
=-2\frac{\kappa\gamma\dot{r}}{c^{2}}\frac{\partial T}{\partial r}+O(\frac{1}{c^{4}}), 
\nonumber\\ && T^{r\psi}_{heat}
=-\frac{\kappa\gamma l}{c^{2}} (\dot{\theta}\frac{\partial T}{\partial r}+\frac{\dot{r}}{r^2}\frac{\partial T}{\partial \theta})+O(\frac{1}{c^{4}}), 
\nonumber\\ && T^{r\varphi}_{heat}
=-\frac{\kappa\gamma l}{c^{2}}(\dot{\phi}\frac{\partial T}{\partial r}+\frac{\dot{r}}{r^2\sin^{2}\theta}\frac{\partial T}{\partial \phi})+O(\frac{1}{c^{4}}), 
\nonumber\\ && T^{\psi\psi}_{heat}
=-2\frac{\kappa\gamma\dot{\theta} l^2}{r^{2}c^{2}} \frac{\partial T}{\partial \theta}+O(\frac{1}{c^{4}}), 
\nonumber\\ && T^{\psi \varphi}_{heat}
=-\frac{\kappa\gamma l^{2}}{c^{2}} [\frac{\dot{\theta}}{\sin^{2}\theta}\frac{\partial T}{\partial \phi}+\dot{\phi}\frac{\partial T}{\partial \theta} +O(\frac{1}{c^{4}}), 
\nonumber\\ && T^{\varphi\varphi}_{heat}
=-2\frac{\kappa\gamma\dot{\phi} l^2}{r^{2}\sin^{2}\theta c^{2}}\frac{\partial T}{\partial \phi}+O(\frac{1}{c^{4}}).
\end{eqnarray} 
\section{Energy density of relativistic fluids}\label{s11} 
An important components of energy momentum tensor is $T^{00}$ which shows the energy density of relativistic fluids. So, as we expect the energy density of relativistic fluids is independent of coordinate system and in the flat metric is given as 
\begin{eqnarray}\label{47} 
&&T^{00}=E=\gamma^2\rho c^2 -p-\frac{1}{c^2}(\lambda\nabla v^{2}\cdot\vec{v} 
\nonumber\\ && \qquad +(\zeta-\frac{2\lambda}{3})v^{2}\nabla\cdot\vec{v}-2\kappa\gamma\nabla T\cdot\vec{v}) . 
\end{eqnarray} 
\section{Energy momentum of non-relativistic fluids}\label{s12} 
To calculate the relations of energy momentum of non-relativistic fluids, we use $v<<c$ limit in the relativistic energy momentum relations. 
\subsection{ Non-relativistic energy momentum of perfect fluid } 
We use the non-relativistic limit in equations (\ref{26}) and (\ref{27}), so, the components of non-relativistic energy momentum tensor of perfect fluid in cylindrical and spherical coordinate systems respectively are 
\begin{eqnarray}\label{48} 
&&T^{00}_{Pf}=\rho c^2 -p,\qquad T^{0r}_{Pf}=\rho c\dot{r},\qquad T^{0\psi}_{Pf}=l\rho c\dot{\theta}, 
\nonumber\\ && T^{0z}_{Pf}=\rho c\dot{z},\qquad 
T^{rr}_{Pf}=\rho \dot{r}^2+p, \qquad T^{r\psi}_{Pf}=l\rho \dot{r}\dot{\theta}, 
\nonumber\\ && T^{rz}_{Pf}=\rho \dot{r}\dot{z}, 
\qquad T^{\psi\psi}_{Pf}=l^2(\rho \dot{\theta}^2+\frac{p}{r^2}), 
\nonumber\\ && T^{\psi z}_{Pf}=l\rho \dot{\theta}\dot{z},\qquad 
T^{zz}_{Pf}=\gamma^2\rho \dot{z}^2+p. 
\end{eqnarray} 
\begin{eqnarray}\label{49} 
&&T^{00}_{Pf}=\rho c^2 -p,\qquad T^{0r}_{Pf}=\rho c\dot{r},\qquad T^{0\psi}_{Pf}=l\rho c\dot{\theta}, 
\nonumber\\ && T^{0\varphi}_{Pf}=l\rho c\dot{\phi},\qquad 
T^{rr}_{Pf}=\rho \dot{r}^2+p, \quad T^{r\psi}_{Pf}=l\rho \dot{r}\dot{\theta}, 
\nonumber\\ && T^{r\varphi}_{Pf}=l\rho \dot{r}\dot{\phi}, 
\qquad T^{\psi\psi}_{Pf}=l^2(\rho \dot{\theta}^2+\frac{p}{r^2}), 
\nonumber\\ && T^{\psi \varphi}_{Pf}=l^2\rho \dot{\theta}\dot{\phi},\qquad 
T^{\varphi\varphi}_{Pf}=l^2(\rho \dot{\phi}^2+\frac{p}{r^2\sin^2\theta}). 
\end{eqnarray} 
\subsection{Non-relativistic shear and bulk tensors}
The non-relativistic shear and bulk tensors can be calculated by using $v<<c$ in the relativistic relations of relativistic shear and bulk tensors . 
\subsubsection{In cylindrical coordinate system} 
In the cylindrical coordinate system the components of the non-relativistic bulk tensor are calculated from equation (\ref{31}) as 
\begin{eqnarray}\label{50} 
&&b^{00}=b^{0r}= b^{0\psi}=b^{0z}=b^{r\psi}=b^{rz}=b^{\psi z}=0, 
\nonumber\\ && b^{rr}=\nabla\cdot\vec{v},\quad 
b^{\psi\psi}=\frac{l^{2}}{r^{2}}\nabla\cdot\vec{v},\quad b^{zz}=\nabla\cdot\vec{v}. 
\end{eqnarray} 
So, the non-relativistic components of shear tensor in the cylindrical coordinate system are obtained from equation (\ref{33}) as 
\begin{eqnarray}\label{51} 
&&\sigma^{00}=\sigma^{0r}=\sigma^{0\psi}=\sigma^{0z}=0,\quad 
\sigma^{rr}=\frac{\partial \dot{r}}{\partial r}-\frac{1}{3}\nabla\cdot\vec{v}, 
\nonumber\\ && \sigma^{r\psi}=\frac{l}{2}[\frac{1}{r^2}\frac{\partial \dot{r}}{\partial \theta}+\frac{\partial \dot{\theta}}{\partial r}],\quad 
\sigma^{rz}=\frac{1}{2}[\frac{\partial \dot{r}}{\partial z}+\frac{\partial \dot{z}}{\partial r}], 
\nonumber\\ &&\sigma^{\psi\psi}=l^{2}[\frac{1}{r^2}\frac{\partial \dot{\theta}}{\partial \theta}+\frac{\dot{r}}{r}] 
-\frac{l^2}{ 3r^2}\nabla\cdot\vec{v}, 
\nonumber\\ && \sigma^{\psi z}=\frac{l}{2}[\frac{\partial \dot{\theta}}{\partial z}+\frac{1}{r^{2}}\frac{\partial \dot{z}}{\partial \theta}], 
\quad\sigma^{zz}=\frac{\partial \dot{z}}{\partial z}-\frac{1}{3}\nabla\cdot\vec{v}. 
\end{eqnarray} 
\subsubsection{In spherical coordinate system} 
In the spherical coordinate system, the non-relativistic bulk tensor are derived from equation (\ref{36}) as 
\begin{eqnarray}\label{52} 
&&b^{00}= b^{r0}=b^{0\psi}=b^{0\varphi}=b^{r\psi}=b^{r\varphi}=b^{\psi \varphi}=0,\quad \nonumber\\ &&b^{rr}=\nabla\cdot\vec{v}, \quad 
b^{\psi\psi}=\frac{l^2}{r^{2}}\nabla\cdot\vec{v}, 
\quad 
b^{\varphi\varphi}=\frac{l^2}{r^{2}sin^{2}\theta}\nabla\cdot\vec{v}. \nonumber\\ 
\end{eqnarray} 
We use the equation (\ref{38}) to calculate the non-relativistic components of shear tensor in the spherical coordinate system 
\begin{eqnarray}\label{53} 
&&\sigma^{00}=\sigma^{0r}=\sigma^{0\psi}=\sigma^{0\varphi}=0,\quad 
\nonumber\\ && \sigma^{rr}=\frac{\partial \dot{r}}{\partial r}-\frac{1}{3}\nabla\cdot\vec{v},\quad 
\sigma^{r\psi}=\frac{l}{2}(\frac{1}{r^2}\frac{\partial \dot{r}}{\partial \theta}+\frac{\partial \dot{\theta}}{\partial r}),\quad 
\nonumber\\ && 
\sigma^{r\varphi}=\frac{l}{2}(\frac{1}{r^{2}\sin^{2}\theta}\frac{\partial \dot{r}}{\partial \phi}+\frac{\partial \dot{\phi}}{\partial r}), 
\nonumber\\ &&\sigma^{\psi\psi}=l^2(\frac{1}{r^2}\frac{\partial \dot{\theta}}{\partial \theta}+\frac{\dot{r}}{r})
-\frac{l^2}{3r^{2}}\nabla\cdot\vec{v},\qquad 
\nonumber\\ &&\sigma^{\psi\varphi}=\frac{l^2}{2r^2}(\frac{1}{\sin^2\theta}\frac{\partial \dot{\theta}}{\partial \phi}+\frac{\partial \dot{\phi}}{\partial \theta}), 
\nonumber\\ && \sigma^{\varphi\varphi}=\frac{l^2}{r^{2}\sin^{2}\theta}(\frac{\partial\dot{\phi}}{\partial \phi}+ \frac{\dot{r}}{r}+\cot\theta\dot{\theta}-\frac{1}{3}\nabla\cdot\vec{v}). 
\end{eqnarray} 
After some calculations, we see that these components with $l=1$ are the same as the components of \cite{mi} for the non-relativistic shear and bulk tensors . 
\subsection{Non-relativistic shear stress viscosity energy momentum }
In this section the components of energy momentum tensor of shear stress viscosity of the non-relativistic fluids are derived with $v<<c$ in relativistic components. 
\subsubsection{In cylindrical coordinate system} 
The non-relativistic components of shear stress viscosity energy momentum tensor are derived after using $v<<c$ in equation (\ref{41}) in the cylindrical coordinate as 
\begin{eqnarray}\label{54} 
&&T^{00}_{vis}=T^{0r}_{vis}=T^{0\psi}_{vis}=T^{0z}_{vis}=0, 
\nonumber\\ && T^{rr}_{vis}=-2\lambda\frac{\partial \dot{r}}{\partial r}-(\zeta-\frac{2\lambda}{3})\nabla\cdot\vec{v}, 
\nonumber\\ && T^{r\psi}_{vis}=-l\lambda(\frac{1}{r^2}\frac{\partial \dot{r}}{\partial \theta}+\frac{\partial \dot{\theta}}{\partial r}), 
\quad T^{rz}_{vis}=-\lambda(\frac{\partial \dot{r}}{\partial z}+\frac{\partial\dot{z}}{\partial r}), 
\nonumber\\ && T^{\psi\psi}_{vis}=-2l^2\lambda(\frac{1}{r^2}\frac{\partial \dot{\theta}}{\partial \theta}+\frac{\dot{r}}{r})-\frac{l^2}{r^2}(\zeta-\frac{2\lambda}{3})\nabla\cdot\vec{v},\qquad\qquad\qquad 
\nonumber\\ && T^{\psi z}_{vis}=-l\lambda( \frac{\partial\dot{\theta}}{\partial z}+\frac{1}{r^2}\frac{\partial\dot{z}}{\partial \theta}),\qquad\qquad\qquad 
\nonumber\\ && 
T^{zz}_{vis}=-2\lambda\frac{\partial \dot{z}}{\partial z}-(\zeta-\frac{2\lambda}{3})\nabla\cdot\vec{v}. 
\end{eqnarray} 
\subsubsection{In spherical coordinate system} 
The non-relativistic components of shear stress viscosity energy momentum tensor in the spherical coordinate system are calculate with $v<<c$ in equation (\ref{42}) 
\begin{eqnarray}\label{55} 
&&T^{00}_{vis}=T^{0r}_{vis}=T^{0\psi}_{vis}=T^{0\varphi}_{vis}=0, 
\nonumber\\ && T^{rr}_{vis}=-2\lambda\frac{\partial \dot{r}}{\partial r}-(\zeta-\frac{2\lambda}{3})\nabla\cdot\vec{v}, 
\nonumber\\ && T^{r\psi}_{vis}=-l\lambda(\frac{1}{r^2}\frac{\partial \dot{r}}{\partial \theta}+\frac{\partial \dot{\theta}}{\partial r}),\nonumber\\ && 
T^{r\varphi}_{vis}=-l\lambda(\frac{1}{r^{2}\sin^{2}\theta}\frac{\partial \dot{r}}{\partial \phi}+\frac{\partial \dot{\phi}}{\partial r}), 
\nonumber\\ && T^{\psi\psi}_{vis}=-\frac{2l^2\lambda}{r^2}(\frac{\partial \dot{\theta}}{\partial \theta}+\frac{\dot{r}}{r})-\frac{l^2}{r^2}(\zeta-\frac{2\lambda}{3})\nabla\cdot\vec{v}, 
\nonumber\\ && T^{\psi\varphi}_{vis}=-l^2\lambda(\frac{1}{r^2\sin^{2}\theta}\frac{\partial \dot{\theta}}{\partial \phi}+\frac{\partial \dot{\phi}}{\partial \theta}), 
\nonumber\\ && T^{\varphi\varphi}_{vis}=-\frac{l^2}{r^{2}\sin^{2}\theta} [ 
2\lambda(\frac{\partial \dot{\phi}}{\partial \phi} 
+\frac{\dot{r}}{r} +\cot\theta\dot{\theta}) 
+(\zeta-\frac{2\lambda}{3})\nabla\cdot\vec{v}].\nonumber\\ 
\end{eqnarray} 
\subsection{Heat flux of the non-relativistic fluids} 
In the non-relativistic fluids, the components of the heat flux are derived by $v<<c$. 
\subsubsection{In cylindrical coordinate system} 
We use the limit of non-relativistic case in equation (\ref{43}). So, the components of the non-relativistic heat flux in the cylindrical coordinate system are 
\begin{equation}\label{56} 
q^{0}=0,\quad q^{r}=-\kappa\frac{\partial T}{\partial r},\quad q^{\psi}=-\frac{\kappa l}{r^{2}}\frac{\partial T}{\partial \theta},\quad q^{z}=-\kappa \frac{\partial T}{\partial z}. 
\end{equation}
\subsubsection{In spherical coordinate system} 
Also, with equation (\ref{44}), the components of the non-relativistic heat flux in the spherical coordinate system are derived as
\begin{equation}\label{57} 
q^{0}=0,\quad q^{r}=-\kappa\frac{\partial T}{\partial r},\quad q^{\psi}=-\frac{\kappa l}{r^{2}}\frac{\partial T}{\partial \theta},\quad q^{\varphi}=-\frac{\kappa l}{r^{2}\sin^{2}\theta}\frac{\partial T}{\partial \phi}. 
\end{equation}
\subsection{Non-relativistic energy momentum of heat flux } 
The energy momentum is created by the heat flux in the non-relativistic fluids can be derived from the relativistic heat flux energy momentum in the limit of $v<<c$. So we see that all components of non-relativistic heat flux energy momentum of cylindrical and spherical coordinate systems are negligiblee.
\subsection{Non relativistic energy density of fluids}
In the non-relativistic fluids, the $T^{00}$ (energy density) is given by 
\begin{equation}\label{58} 
T^{00}=E=\rho c^2 -p.
\end{equation} 
\section{Summery and Conclusion}\label{s13} 
In this paper, the components of energy momentum tensor and its elements are studied in the flat metric, with the ignorable magnetic field. So, there are three types of energy momentum tensors, which are perfect fluid, shear stress viscosity and heat flux energy momentum tensors.
 
The iso-dimension scale is presented in this paper. In this scale, the $c=1$ is not used and all components of metric are dimensionless, also for a sample variable, all components have the same dimension. All calculations are done in this scale. So, if we choose a suitable $l$, we can compare the different components of a variables, for example in the energy momentum tensor, it is seen that which components or elements of energy momentum are more important.

In this paper, the components of bulk, shear, shear stress viscosity energy momentum, heat flux and heat flux energy momentum tensors are calculated in cylindrical and spherical coordinate systems.

The iso-dimension scale makes connection between relativistic and non-relativistic quantities. So, the components of bulk, shear, shear stress viscosity energy momentum, heat flux and heat flux energy momentum tensors of non-relativistic fluids are calculated with the relativistic ones. Also, some the differences of non-relativistic and relativistic fluids are seen in this scale. This connection enables us to use some corrections in the studies of non-relativistic fluids to give the more accurate results. For example, in the first order corrections, the $\frac{v}{c}$ terms of shear, bulk, heat flux and energy momentum tensors are used.

The shear and shear stress viscosity energy momentum tensors of relativistic fluids have 16 non-zero components, but the shear and shear stress viscosity energy momentum tensors of non-relativistic fluids have 9 non-zero components. Relativistic heat flux energy momentum has 16 non-zero components but all components of non-relativistic heat flux energy momentum are dispensable. 
The non-relativistic bulk tensor just have three diagonal components which is proportional to $\nabla\cdot\vec{v}$, but in the relativistic bulk tensor 16 components are seen, also, time derivative of velocity has influences in the diagonal spatial components. In the non- relativistic fluids, shear and shear stress viscosity energy momentum tensors is generated by spatial derivatives of velocity, but in the relativistic fluids, time and spatial derivatives of velocity create shear and shear stress viscosity energy momentum tensors.

In the non-relativistic fluids, heat flux is generated with the spatial derivatives of temperature, so, $i$ (i is for spatial components) component is proportional to $\frac{\partial T}{\partial x^{i}}$. But the relativistic heat flux is generated with the time and spatial derivatives of velocity and temperature. So, in the $i$ component of heat flux, all components of $\frac{\partial T}{\partial x^{j}}$ is seen. Specially we see that, in the relativistic fluids the heat flux is generated in a constant temperature.

In the relativistic fluids, the energy density is the same in two coordinate systems. In the energy density of viscosity the important terms are generated by the spatial derivative of velocity, also the main terms of heat flux energy density are produced by the gradient of velocity, so the time derivatives of velocity, temperature, etc are not important in the energy density. Energy density includes influences of perfect fluid, shear stress viscosity and heat flux, but, in the non-relativistic fluids, the influences of shear stress viscosity and heat flux are not seen in the energy. 
\section{Some examples and best-practices}
\label{sec:intro}

\appendix
\section{Projection tensor } 
The components of projection tensor are derived from equation (\ref{8}) in cylindrical and spherical coordinate systems. 
\subsection{In the cylindrical coordinate system} 
The components of projection tensor $h^{\mu\nu}$ in the iso-dimension sale coordinate are shown as 
\begin{eqnarray}\label{a1} 
&&h^{00}=-1+\gamma^{2}= \frac{v^{2}}{c^{2}}+O(\frac{1}{c^{4}}),\qquad h^{0r}=\frac{\gamma^{2}\dot{r}}{c}=\frac{\dot{r}}{c}(1+\frac{v^{2}}{c^{2}})+O(\frac{1}{c^{5}}), \nonumber\\
&&h^{0\psi}=\frac{l\gamma^{2}\dot{\theta}}{c}=\frac{l\dot{\theta}}{c}(1+\frac{v^{2}}{c^{2}})+O(\frac{1}{c^{5}}),\,
h^{0z}=\frac{\gamma^{2}\dot{z}}{c}=\frac{\dot{z}}{c}(1+\frac{v^{2}}{c^{2}})+O(\frac{1}{c^{5}}), \nonumber\\ &&h^{rr}=1+\frac{\gamma^{2}\dot{r}^{2}}{c^{2}}=1+\frac{\dot{r}^{2}}{c^{2}}+O(\frac{1}{c^{4}}),\qquad 
h^{\psi r}=\frac{l\gamma^{2}\dot{r}\dot{\theta}}{c^{2}}=\frac{l\dot{r}\dot{\theta}}{c^{2}}+O(\frac{1}{c^{4}}),\nonumber\\
&&h^{rz}=\frac{\gamma^{2}\dot{r}\dot{z}}{c^{2}}=\frac{\dot{r}\dot{z}}{c^{2}}+O(\frac{1}{c^{4}}),\,
h^{\psi\psi}=\frac{l^2}{r^2}+\frac{l^2\gamma^{2}\dot{\theta}^{2}}{c^{2}}=\frac{l^2}{r^2}+\frac{l^2\dot{\theta}^{2}}{c^{2}}+O(\frac{1}{c^{4}}),\nonumber\\
&& h^{\psi z}=\frac{l\gamma^{2}\dot{\theta}\dot{z}}{c^{2}}=\frac{l\dot{\theta}\dot{z}}{c^{2}}+O(\frac{1}{c^{4}}),\quad 
h^{zz}=1+\frac{\gamma^{2}\dot{z}^{2}}{c^{2}}=1+\frac{\dot{z}^{2}}{c^{2}}+O(\frac{1}{c^{4}}). \nonumber\\
\end{eqnarray} 
\subsection{In the spherical coordinate system} 
The components of projection tensor in the iso-dimension sale coordinate in the spherical coordinate system are 
\begin{eqnarray}\label{a2} 
&&h^{00}= \frac{v^{2}}{c^{2}}+O(\frac{1}{c^{4}}) ,\qquad h^{0r}=\frac{\dot{r}}{c}(1+\frac{v^{2}}{c^{2}})+O(\frac{1}{c^{5}}), 
\nonumber\\ &&h^{t\psi}=\frac{l\dot{\theta}}{c}(1+\frac{v^{2}}{c^{2}})+O(\frac{1}{c^{5}}),\quad 
h^{t\varphi}=\frac{l\dot{\phi}}{c}(1+\frac{v^{2}}{c^{2}})+O(\frac{1}{c^{5}}), 
\nonumber\\ &&h^{rr}=1+\frac{\dot{r}^{2}}{c^{2}},\qquad 
h^{r\psi}=\frac{l\dot{r}\dot{\theta}}{c^{2}}+O(\frac{1}{c^{4}}), 
\nonumber\\ &&h^{r\varphi}=\frac{l\dot{r}\dot{\phi}}{c^{2}}+O(\frac{1}{c^{4}}),\qquad 
h^{\psi\psi}=\frac{l^2}{r^2}+\frac{l^2\dot{\theta}^{2}}{c^{2}}+O(\frac{1}{c^{4}}), 
\nonumber\\ && h^{\psi\varphi}=\frac{l^2\dot{\theta}\dot{\phi}}{c^{2}}+O(\frac{1}{c^{4}})\quad 
h^{\varphi\varphi}=\frac{l^2}{r^2 sin^{2}\theta}+\frac{l^2\dot{\phi}^{2}}{c^{2}}+O(\frac{1}{c^{4}}). \nonumber\\ 
\end{eqnarray} 

\acknowledgments
This research was supported by the grant of Kosar
university of Bojnord with the grant number of
”NO.9610171345”.
\\\textbf{Data availabity statment}\\
The data underlying this article are available in the article.


\end{document}